\documentclass[12pt]{article}
\usepackage{amssymb}
\usepackage{amsmath}
\usepackage{amssymb,amsmath,cite,epsfig}
\usepackage[top=2.5cm, bottom=2.5cm, left=2.5cm, right=2.5cm]{geometry}
\usepackage{indentfirst}
\usepackage{subfig}
\usepackage{float}
\usepackage{placeins}

\begin{document}

\renewcommand*{\thefootnote}{\fnsymbol{footnote}}

\begin{titlepage}
\vspace*{2cm}
\begin{center}
{\Large\textsf{\textbf{Particle Creation in Cosmological Space-time\\ by a Time-Varying Electric Field}}}
\end{center}
\par \vskip 5mm
\begin{center}
{\large\textsf{Hadjar Rezki\textsuperscript{a,b} and Slimane Zaim\textsuperscript{b}\footnote{Corresponding author.}}}\\
\vskip 5mm
\textsuperscript{a} Laboratory of Physics of Radiation and its Interactions with Matter, \\
Department of Physics, Faculty of Matter Sciences,\\
University of Batna 1, Batna 05000, Algeria \\
\textsuperscript{b} Department of Physics, Faculty of Matter Sciences,\\
University of Batna 1, Batna 05000, Algeria \\
\end{center}
\par \vskip 2mm
\begin{center}
E-mail: hadjar.rezki@univ-batna.dz, zaim69slimane@yahoo.com\\
\end{center}
\par \vskip 2mm
\begin{center} {\large \textsf{\textbf{Abstract}}}\end{center}
\begin{quote}
\quad
In this work, we employ the semi-classical approach to solve the Klein-Gordon and Dirac equations in the presence of a time-varying electric field. Our objective is to calculate the density of particle creation in a cosmological anisotropic Bianchi I space-time. We demonstrate that when the electric interaction is proportional to the Ricci scalar of curved space-time, the particle distribution subjected to the electric field transforms into a thermal state.
\end{quote}

\noindent\textbf{\sc Pacs numbers}: 11.10.Nx, 03.65.Pm, 03.70.+k , 25.75.Dw

\end{titlepage}

\renewcommand*{\thefootnote}{\arabic{footnote}}

\section{Introduction}

Black holes, according to classical theory, are believed to solely absorb particles and not emit them. However, quantum mechanical phenomena have demonstrated that black holes can indeed produce and emit particles. The significance of particle production in cosmology and astrophysics is widely acknowledged. Pioneering research conducted by Parker and others \cite{1,2,3,4} has extensively explored particle creation in the early universe driven by gravitational field effects. Consequently, the investigation of quantum effects in cosmology has become a highly desirable research area. Surprisingly, there have been limited attempts to address these effects in anisotropic universes \cite{1,2,3,4}.

The phenomenon of particle creation from the vacuum under the influence of strong external electromagnetic and gravitational fields has been a subject of study for a considerable duration, as evidenced by references~\cite{5,6,7,8,9,10,11,12,13,14,15,16,17,18}. Schwinger initiated the exploration of particle creation effects long ago~\cite{6}. These effects, coupled with back-reaction considerations, are crucial in understanding black hole physics. Recent advancements in laser physics raise the prospect of reaching the non-perturbative regime of pair production in the near future~\cite{19}. In laboratory conditions, electron-hole pair creation from the vacuum has become observable in graphene and similar nanostructures~\cite{20,21,22,23}.

Depending on the strong field structure, various approaches have been proposed to calculate non-perturbative effects. The semiclassical approximation, like any other approach, is formulated within the framework of quantum field theory. Particle production is heavily dependent on the space-time structure of the external field, leading to expectations of non-locality, final-state correlations, and real-time dynamics in the pair production process.

The role of particle creation from the vacuum by external electric and gravitational backgrounds is paramount in cosmology and astrophysics~\cite{24,25}. Despite this, there are only a few exactly solvable cases known for either time-dependent homogeneous or constant inhomogeneous electric fields. Notable examples include constant uniform electric fields~\cite{1}, adiabatic electric fields~\cite{26,27}, and $T$-constant electric fields that turn on and off at definite time instants $t_{in}$ and $t_{fin}$, with $t_{fin}-t_{in}=T$ being constant within the time interval $T$. Other cases involve periodic alternating electric fields~\cite{30,31}, exponentially growing and decaying electric fields~\cite{29,32,33,34}, and various constant inhomogeneous electric fields where time, $t$, is replaced by the spatial coordinate, $x$. Some electric fields satisfying more complicated symmetries, such as those with potentials given in light-cone variables~\cite{35,36,37}, and an inverse-square electric field (an electric field inversely proportional to the square of time)~\cite{38,39}, have also been explored.

In this work, we specifically investigate vacuum instability in an electric field linearly proportional to the inverse of time plus the inverse of the square of time. This behavior is characteristic of an effective mean electric field in graphene, resulting from a second-order deformation of the initial constant electric field due to backreaction caused by vacuum instability~\cite{40}.

Various techniques are available for studying particle creation in cosmological backgrounds. These include the adiabatic approach~\cite{1,41}, the Feynman path integral method~\cite{42}, the Hamiltonian diagonalization technique~\cite{43,44}, and the Green function approach~\cite{45}.

In this research paper, the semi-classical method is employed to investigate the well-posed cosmological problem of particle creation by a non-stationary external electric field in a non-static space-time. The semi-classical method involves a multi-step process. First, the relativistic Hamilton-Jacobi equation is solved. By examining its solutions as $t\rightarrow 0$ and $t\rightarrow \infty$, positive and negative frequency modes are identified. Second, the Klein-Gordon and Dirac equations are solved. By comparing the results with those obtained for the quasi-classical limit, positive and negative frequency states are identified for large values of time ($t\rightarrow \infty$) and in the vicinity of the initial singularity ($t\rightarrow 0$). These modes correspond to positive and negative eigenvalues, respectively. In the third step, the number density of created particles is computed. This is achieved using Bogoliubov transformations, establishing a connection between negative frequency states for large values of time ($t\rightarrow \infty$) and positive and negative frequency states in the vicinity of the initial singularity ($t\rightarrow 0$).

Due to the limited availability of simple models, addressing particle creation in time-dependent electromagnetic fields can pose technical challenges~\cite{43}. An intriguing scenario necessitating a discussion on particle creation arises in the presence of a non-stationary electric field in a Bianchi I anisotropic universe, defined by the metric:
\begin{equation}\label{eq:1}
ds^{2}=-dt^{2}+t^{2}\left( dx^{2}+dy^{2}\right) +t^{-2}dz^{2}\,.
\end{equation}
Determining particle states using the adiabatic approach is challenging due to the metric's space-like singularity at $t=0$~\cite{1,41}. Hence, we employ the quasi-classical method to tackle this issue.

In an anisotropic Bianchi I universe with both constant and time-varying electric fields, previous research has explored the creation of scalar and Dirac particles~\cite{39,46,47,48,49,50}. These studies reveal that overlooking electrical interaction results in a thermal particle distribution. Our objective is to investigate the production of scalar and Dirac particles from vacuum in a non-stationary electric field within the context of an anisotropic Bianchi I spacetime. Our findings indicate that when both anisotropic and non-stationary electric fields are present, particles are produced in a manner such that the distribution density becomes thermal when the electric field is proportional to the Ricci scalar curvature $R$.

This paper is organized as follows. In Section 2, we solve the relativistic Hamilton-Jacobi equation and compute the quasi-classical energy modes. Section 3 is dedicated to solving the Klein-Gordon equation, followed by the computation of the density of created scalar particles and a discussion on weak and strong field limits. In Section 4, we extend our analysis to solve the Dirac equation and calculate the density of created Dirac particles. The paper concludes with a discussion in the final section.

\section{Hamilton Jacobi equation}

The main principle of this approach is as follows. To determine the positive and negative frequency states, we first solve the Hamilton-Jacobi equation, the Klein-Gordon equation, and the Dirac equation for the Bianchi I space-time in the presence of a time-dependent electromagnetic field. Subsequently, we compare the asymptotic behavior of the solutions of the field equations with the Hamilton-Jacobi equation. The positive and negative frequency states are obtained using the Hamilton-Jacobi method in accordance with the following expressions
\begin{equation}
\varphi\left(\vec{x},t\right)=\exp\left(-iS\right),\qquad\text{for positive frequency,}
\end{equation}
and
\begin{equation}
\varphi\left(\vec{x},t\right)=\exp\left(iS\right),\qquad\text{for negative frequency,}
\end{equation}
where $S$ stands for the classical action.

In the presence of an external electromagnetic field, represented by potentials $A_{\mu}$, the Hamilton-Jacobi equation for a relativistic particle with rest mass $m$ and electric charge $e$ is expressed as
\begin{equation}\label{eq:4}
g^{\mu\nu}\left(\frac{\partial S}{\partial x^{\mu}}+eA_{\mu }\right)\left(\frac{\partial S}{\partial x^{\nu}}+eA_{\nu}\right)+m^{2}=0\,,
\end{equation}
where $g^{\mu\nu}$ is the general inverse matrix of the tensor matrix $g_{\mu\nu}$, defined by the tetrad fields $e_{\mu}^{b}$ (which are a set of four basis vectors for space-time) according to the form
\begin{equation}
g_{\mu\nu}=e_{\mu}^{b}e_{\nu}^{a}\eta_{ba}\,,\qquad g^{\mu\nu}=e_{b}^{\mu}e_{a}^{\mu}\eta^{ba}\,,
\end{equation}
where $e$ represents the electric charge, and $A_{\mu}$ is the electromagnetic four-potential. It is worth noting that, for simplicity, we adopt the conventions of $c=1$ and $\hbar=1$ throughout this context and elsewhere.

Considering the cosmic scenario described by the line element in equation \eqref{eq:1}, the metric tensor matrix $g$ can be diagonalized as follows:
\begin{equation}\label{eq:6}
g_{\mu\nu}=\left(
\begin{array}{cccc}
-1 & 0 & 0 & 0 \\
0 & t^{2} & 0 & 0 \\
0 & 0 & t^{2} & 0 \\
0 & 0 & 0 & t^{-2}
\end{array}
\right),
\end{equation}
with the corresponding inverse matrix  $g^{\mu\nu}$
\begin{equation}
g^{\mu \nu }=\left(
\begin{array}{cccc}
-1 & 0 & 0 & 0 \\
0 & t^{-2} & 0 & 0 \\
0 & 0 & t^{-2} & 0 \\
0 & 0 & 0 & t^{2}
\end{array}
\right).
\end{equation}
For further simplifications, we adopt a diagonal tetrad
\begin{align}
e_{\mu }^{0}& =\left(\begin{array}{cccc}1, & 0, & 0, & 0\end{array}\right),\\
e_{\mu }^{1}& =\left(\begin{array}{cccc}0, & t, & 0, & 0\end{array}\right),\\
e_{\mu }^{2}& =\left(\begin{array}{cccc}0, & 0, & t, & 0\end{array}\right),\\
e_{\mu }^{3}& =\left(\begin{array}{cccc}0, & 0, & 0, & t^{-1}\end{array}\right).
\end{align}

Moreover, we introduce a time-dependent external four-vector field $A_{\mu}$, which activates at $t=0$, deactivates at the infinitely remote future $t=+\infty$, and is linearly proportional to the inverse of time plus the inverse of the square of time. The field is uniformly distributed over space and directed along the $z$-axis. The four-vector $A_{\mu}$ is chosen such that:
\begin{equation}\label{eq:12}
A_{\mu} =E\left(0,0,0,\frac{\tau_{1}}{t}+\frac{\tau_{2}^{2}}{t^{2}}\right).
\end{equation}
In the vicinity of the $t=0$, the electric field diverges. But, by virtue of a shift in the time coordinate, the point of divergence can be shifted to any arbitrary instant in time. The electric field $(12)$ is parameterized by two constants $\tau_{1,2}$, which play the role of time scales for the pulse durations, respectively. This is equivalent to a variable electric field in the $z$-direction, evidenced by the presence of invariants
\begin{equation}
F_{\mu\nu}F^{\mu\nu}\neq0\,,\quad\text{and}F_{\mu\nu}^{\ast}F^{\mu\nu}=0\,.
\end{equation}
We note that there is no magnetic field here. Thus, the matrices of $g^{\mu \nu}$ and $A_{\mu}$ depend only on time, and the $S$ function can be divided as follows
\begin{equation}\label{eq:14}
S=\vec{k}\cdot\vec{r}+F\left(t\right).
\end{equation}
Substituting eq. \eqref{eq:14} into eq. \eqref{eq:4}, we determine the equation for $F(t)$
\begin{equation}\label{eq:15}
-\left(\frac{dF}{dt}\right)^{2}+\frac{k_{\perp }^{2}}{t^{2}}+t^{2}k_{z}^{2}+\left(ik_{z}t+e\alpha+\frac{e\beta}{t}\right)^{2}+m^{2}=0\,,
\end{equation}
where $\alpha=E\tau_{1}$, $\beta=E\tau_{2}^{2}$, and $k_{\perp}^{2}=k_{x}^{2} + k_{y}^{2}$. Taking the propagation plane of the wave function perpendicular to the direction of the axis of the external electric field, i.e., $k_{z}=0$ and $t\rightarrow 0$, then the solution of equation \eqref{eq:15} is
\begin{equation}
F\left(t\right)=\pm\sqrt{k_{\perp }^{2}+e^{2}\beta^{2}}\log t\,.
\end{equation}
Hence, the wave function $\varphi(\vec{r},t)$ reads
\begin{equation}
\varphi\left(\vec{r},t\right)\sim\exp\left(i\vec{k}\cdot\vec{r}\right)t^{\pm\sqrt{k_{\perp}^{2}+e^{2}\beta^{2}}}\,.
\end{equation}
For $k_{z}=0$ and $t\rightarrow\infty$, the solution of equation \eqref{eq:15} is given by
\begin{equation}
F\left(t\right)=\pm\sqrt{m^{2}+e^{2}\alpha^{2}}t\,.
\end{equation}
Hence, the wave function $\varphi(\vec{r},t)$ reads
\begin{equation}
\varphi\left(\vec{r},t\right)\sim\exp\left(i\vec{k}\cdot \vec{r}\right)\exp\left(\pm\sqrt{m^{2}+e^{2}\alpha^{2}}t\right).
\end{equation}

The Klein-Gordon and Dirac equations must be solved in Bianchi I space-time in the presence of a time-dependent electric field to find the negative and positive frequency states of the scalar and Dirac particles.

\section{Klein-Gordon equation and particle creation process}

In the presence of an electrodynamic gauge field $A_{\mu}$ within curved space-time, the simplest action governing a massive scalar field takes the form:
\begin{equation}\label{eq:20}
\mathcal{S}=\frac{1}{2}\int d^4x\,\Xi\left(g^{\mu\nu}\left(D_{\mu}\varphi\right)^\dagger D_{\nu}\varphi+\zeta R\varphi^\dagger\varphi+m^2\varphi^\dagger\varphi\right),
\end{equation}
where $\Xi=\det e_{\mu}^{b}$ serves as the Jacobian to ensure the action remains a scalar under general coordinate transformations. Here, $D_{\mu}\varphi=\left(\partial_{\mu}-ieA_{\mu}\right)\varphi$ denotes the gauge covariant derivative. The Ricci scalar $R$ is defined as
\begin{equation}
R=g^{\mu\nu}R_{\mu\nu}\,,
\end{equation}
with the Ricci tensor $R_{\mu \nu }$ given by
\begin{equation}
R_{\mu\nu}=\partial_\alpha\Gamma_{\mu\nu}^{\alpha}-\partial_{\nu}\Gamma_{\mu\alpha}^{\alpha}+\Gamma_{\beta\alpha}^{\alpha}\Gamma_{\mu\nu}^{\beta}-\Gamma_{\mu\beta}^{\alpha}\Gamma_{\alpha\nu}^{\beta}
\end{equation}
where $\Gamma_{\mu\nu}^{\alpha}$ represents the Christoffel symbols defined as
\begin{equation}
\Gamma_{\mu\nu}^{\alpha}=\frac{1}{2}g^{\alpha\gamma}\left(\partial_{\nu}g_{\mu\gamma}+\partial_{\mu}g_{\gamma\nu}-\partial_{\gamma}g_{\mu\nu}\right),
\end{equation}

Utilizing infinitesimal generic field transformations ($\delta_{\lambda}\varphi =i\lambda\varphi$), we can demonstrate the invariance of the action in equation \eqref{eq:20}. By applying the Noether theorem and varying the scalar density under the gauge transformation, we obtain the generalized field equation:
\begin{equation}
\frac{\partial\mathcal{L}}{\partial\varphi}-\partial_{\mu}\frac{\partial\mathcal{L}}{\partial\left(\partial_{\mu}\varphi\right)}=0\,.
\label{eq:field}
\end{equation}
Consequently, the Klein-Gordon equation in curved space with the presence of an external field is expressed as
\begin{equation}\label{eq:25}
\left(-\square +m^2+\zeta R\right)\varphi =0\,,
\end{equation}
where
\begin{equation}
\square =\frac{1}{\Xi}D_{\mu}\left(\Xi e_{a}^{\mu }e^{\nu a}D_{\nu}\right),
\end{equation}
and the scalar curvature $R$ is given by
\begin{equation}
R=2/t^{2}\,.
\end{equation}
The determinant of the corresponding metric tensor $g_{\mu \nu }$ in equation \eqref{eq:6} is
\begin{equation}
\Xi =\sqrt{\left\vert \det g\right\vert }=t\,.
\end{equation}

In the context of the adiabatic approach \cite{1,2,3,4,41}, its application is precluded for identifying particle states due to a singularity at $t=0$ in the line element \eqref{eq:1}. The standard method for solving the Klein-Gordon equation involves a comparison with quasi-classical limits to specify positive and negative frequency states. Subsequently, through Bogoliubov transformations, one can deduce the number density of created particles.

The Klein-Gordon equation \eqref{eq:25} can be expressed in a simplified form
\begin{equation}\label{eq:29}
\left(g^{\mu \nu}\partial_{\mu}\partial_{\nu}-m^{2}-\xi R\right)\varphi-\left(ieg^{\mu\nu}\partial_{\mu}A_{\nu}+e^{2}g^{\mu\nu}A_{\mu}A_{\nu}+2ieg^{\mu\nu}A_{\mu}\partial_{\nu}\right)\varphi=0\,,
\end{equation}
Here, the terms are further detailed
\begin{equation}
g^{\mu\nu}\partial_{\mu}\partial_{\nu}=-\partial_{0}^{2}-\frac{1}{t}\partial_{0}+\frac{1}{t^{2}}\left(\partial_{1}^{2}+\partial_{2}^{2}\right)+t^{2}\partial_{3}^{2}\,,
\end{equation}
and
\begin{equation}
2ieg^{\mu\nu}A_{\mu}\partial_{\nu}=2ie\left(\alpha t+\beta\right)\partial_{3}\,,
\end{equation}
and
\begin{equation}
e^{2}g^{\mu\nu }A_{\mu }A_{\nu}=\left[e\left(\alpha+\frac{\beta}{t}\right)\right]^{2}.
\end{equation}
Subsequently, the Klein-Gordon equation \eqref{eq:29} simplifies to
\begin{equation}\label{eq:33}
\left[-\partial_{0}^{2}-\frac{1}{t}\partial_{0}+\frac{1}{t^{2}}\left(\partial_{1}^{2}+\partial_{2}^{2}\right)+t^{2}\partial_{3}^{2}-m^{2}-\xi R-2ie\left(\alpha t+\beta\right)\partial_{3}-e^{2}\left(\alpha
+\frac{\beta}{t}\right)^{2}\right]\varphi =0\,.
\end{equation}
Notably, this equation commutes with the operator $-i\overrightarrow{\nabla}$, allowing the wave function $\hat{\varphi}$ to be expressed as
\begin{equation}\label{eq:34}
\varphi=t^{-1/2}w(t)\exp\left(ik_{x}x+ik_{y}y+ik_{z}z\right).
\end{equation}
By substituting equation \eqref{eq:34} into equation \eqref{eq:33}, one obtains
\begin{equation}\label{eq:35}
\left[\frac{d^{2}}{dt^{2}}+\frac{k_{\bot}^{2}+e^{2}\beta^{2}+\frac{1}{4}+\xi}{t^{2}}+2\frac{e^{2}\alpha\beta}{t}+t^{2}k_{z}^{2}-2e\left(\alpha t+\beta\right)k_{z}+m^{2}+e^{2}\alpha^{2}\right]w(t)=0\,,
\end{equation}
where the eigenvalue $k_{\bot}$ is defined as
\begin{equation}
k_{\bot}=\sqrt{k_{x}^{2}+k_{y}^{2}}\,.
\end{equation}
Introducing the variable change
\begin{equation}
\rho=-2i\sqrt{m^{2}+e^{2}\alpha^{2}}t\,,
\end{equation}
we conclude that for $k_{z}=0$ equation \eqref{eq:35} transforms to
\begin{equation}\label{eq:38}
\left[\frac{d^{2}}{d\rho^{2}}+\frac{\frac{1}{4}-\mu^{2}}{\rho^{2}}+\frac{\lambda}{\rho}-\frac{1}{4}\right]w(\rho)=0\,.
\end{equation}
where
\begin{equation}
\mu=i\sqrt{k_{\bot}^{2}+e^{2}\beta^{2}+2\xi}\,,
\end{equation}
and
\begin{equation}
\lambda=i\frac{e^{2}\alpha\beta}{\sqrt{m^{2}+e^{2}\alpha^{2}}}\,.
\end{equation}

Equation \eqref{eq:38} corresponds to the Whittaker differential equation \cite{51,52,53}. Consequently, the solutions can be expressed as a combination of Whittaker functions, denoted as $M_{\tilde{k}_{\theta},\mu}\left( \rho\right) $ and $W_{\tilde{k}_{\theta },\mu }\left( \rho \right)$
\begin{equation}
w\left(\rho\right)=C_{1}M_{\lambda,\mu}\left(\rho\right)+C_{2}W_{\lambda,\mu}\left(\rho \right),\label{eq:sol}
\end{equation}
where $C_{1}$ and $C_{2}$ represent normalization constants. The linear Whittaker functions $M_{\lambda,\mu}\left(\rho\right)$ are defined as
\begin{equation}
M_{\lambda,\mu}\left(\rho\right)=\rho^{\mu+1/4}e^{-\rho /2}\left[1+\frac{\frac{1}{2}+\mu-\lambda}{1!\left(2\mu+1\right)}\rho+\frac{\left(\frac{1}{2}+\mu-\lambda\right)
\left(\frac{3}{2}+\mu-\lambda \right)}{2!\left(2\mu+1\right)\left(2\mu+2\right)}\rho^{2}+\cdots\right],
\end{equation}
and
\begin{equation}\label{eq:43}
W_{\lambda,{\mu}}\left(\rho\right)=\frac{\Gamma\left(-2\mu\right)}{\Gamma\left(\frac{1}{2}-\mu-\lambda\right)}M_{\lambda,\mu}\left( \rho \right) +\frac{\Gamma \left(2\mu\right)}{\Gamma
\left(\frac{1}{2}+\mu -\lambda\right)}M_{\lambda ,-{\mu }}\left(\rho \right).
\end{equation}

It is worth noting that the Whittaker functions $M_{\lambda,\mu}\left(\rho\right)$ and $W_{\lambda,\mu}\left(\rho\right)$ exhibit the following asymptotic behavior
\begin{eqnarray}
M_{\lambda,\mu}\left(\rho\right)&\sim&\rho^{\mu+1/2}\,, \\
W_{\lambda,\mu}\left(\rho\right)&\sim&\frac{\Gamma\left(2\mu\right)}{\Gamma\left(\frac{1}{2}+\mu-\lambda\right)}\rho^{-\mu+1/2}\,,
\end{eqnarray}
for $\rho \rightarrow 0$, and
\begin{eqnarray}
M_{\lambda,\mu}\left(\rho\right)&\sim&\frac{\Gamma\left(2\mu\right)}{\Gamma\left(\frac{1}{2}+\mu-\lambda\right)}\rho^{-\lambda}\exp\left(\rho/2\right),\\
W_{\lambda,{\mu}}\left(\rho\right)&\sim&\rho^{\lambda }\exp\left( -\rho /2\right),
\end{eqnarray}
and for $\rho\rightarrow\infty$.

By utilizing the asymptotic limits of the solution \eqref{eq:38} at $t=0$ ($\rho=0$) and $t\rightarrow\infty$, we can construct the positive and negative frequency modes. Consequently, the positive and negative frequency modes take the form
\begin{equation}
w_{0}^{+}=C_{0}^{+}M_{\lambda,\mu}\left(\rho\right),
\end{equation}
and
\begin{equation}
w_{0}^{-}=\left(w_{0}^{+}\right)^{\ast}=C_{0}^{+}\left(-1\right)^{-\mu+\frac{1}{2}}M_{\lambda,-{\mu}}\left(\rho\right),
\end{equation}
It is straightforward to find the positive and negative frequency modes for $\left\vert\rho\right\vert\rightarrow\infty$ by examining the asymptotic limit of $W_{\lambda,\mu}\left(\rho\right)$, yielding:
\begin{equation}
w_{\infty}^{+}=C_{\infty}^{+}W_{\lambda,\mu}\left(\rho\right),
\end{equation}
and
\begin{equation}
w_{\infty}^{-}=C_{\infty}^{-}W_{-\lambda,\mu}\left(-\rho\right),
\end{equation}
where $C_{\infty}^{+}$ and $C_{\infty}^{-}$ are normalization constants.

Through the Bogoliubov transformation, we can express the positive frequency mode $w_{\infty}^{+}$ in terms of the positive ($w_{0}^{+}$) and negative ($w_{0}^{-}$) frequency modes
\begin{equation}\label{eq:52}
w_{\infty}^{+}=\alpha w_{0}^{+}+\beta w_{0}^{-}\,,
\end{equation}
According to relation \eqref{eq:43}, equation \eqref{eq:52} can also be rewritten as follows
\begin{align}
&C_{\infty }^{+}\,\frac{\Gamma\left(-2\mu \right)}{\Gamma\left(\frac{1}{2}-\mu-\lambda\right)}M_{\lambda,\mu}\left(\rho\right)+C_{\infty}^{+}\,\frac{\Gamma\left(2\mu\right)}
{\Gamma \left(\frac{1}{2}+\mu-\lambda\right)}M_{\lambda,-\mu}\left(\rho\right)\notag \\
&=\alpha\,C_{0}^{+}\,M_{\lambda,\mu}\left(\rho\right)+\beta\,C_{0}^{+}\left(-1\right)^{-\mu+\frac{1}{2}}M_{\lambda,-\mu}\left( \rho \right).
\end{align}

To find $\alpha$ and $\beta$, we use the expressions:
\begin{equation}
\alpha=\frac{C_{\infty }^{+}\Gamma \left( -2\mu \right) }{C_{0}^{+}\Gamma\left( \frac{1}{2}-\lambda -\mu \right) },\qquad \beta =\frac{C_{\infty}^{+}\Gamma \left( 2\mu \right) }{C_{0}^{+}\Gamma \left( \frac{1}{2}-\lambda
+\mu \right) }\exp \left( i\pi \left( \mu -1/2\right) \right),
\end{equation}
with
\begin{equation}
\frac{\left\vert \beta \right\vert ^{2}}{\left\vert \alpha \right\vert ^{2}}=\left\vert \frac{\Gamma \left( \frac{1}{2}-\lambda -\mu \right) }{\Gamma\left( \frac{1}{2}-\lambda +\mu \right) }\right\vert ^{2}\exp \left( -2\pi
\mu \right) .
\end{equation}
Utilizing the property of the Gamma function
\begin{equation}
\left\vert\Gamma\left( \frac{1}{2}+i\rho \right) \right\vert ^{2}=\frac{\pi }{\cosh \left( \pi \rho \right) },
\end{equation}
and simplifying, we arrive at
\begin{equation}
\frac{\left\vert \beta \right\vert ^{2}}{\left\vert \alpha \right\vert ^{2}}=\frac{1+\exp \left( -2\pi \left( \mu -\lambda \right) \right) }{1+\exp\left( 2\pi \left( \mu +\lambda \right) \right) }.
\end{equation}
Subsequently, the probability of producing a single particle in mode $k$ from a vacuum is given by
\begin{equation}
P_{k}=\frac{\left\vert \beta \right\vert ^{2}}{\left\vert \alpha \right\vert^{2}}=\frac{\cosh \left[ \pi \left( -\lambda +\mu \right) \right] }{\cosh \left[ \pi \left( \lambda +\mu \right) \right] }\exp \left( -2\pi
\mu \right) .
\end{equation}

To calculate the created particle density $n_{k}$ due to the effect of the time-dependent electric field, we use equation \eqref{eq:52}, where
\begin{equation}
n_{k}=\left\vert\beta\right\vert^2\,.
\end{equation}
The Bogoliubov coefficients satisfy the relation
\begin{equation}
\left\vert\alpha\right\vert^2-\left\vert\beta\right\vert^2=1\,,
\end{equation}
and the number density of the created particles is given by
\begin{equation}\label{eq:61}
n_{k}=\frac{P_{k}}{1-P_{k}}=\exp\left[-\pi\left(\lambda+\mu\right)\right]\frac{\cosh\left[\pi\left(-\lambda+\mu\right)\right]}{\sinh\left(2\pi\mu\right)}\,.
\end{equation}
\begin{figure}[ht]
\centering
\subfloat[\{$\alpha, -0.01, 0.01$\},\{ $\beta, -1, 1$\}]
{\includegraphics[width=0.35\textwidth]{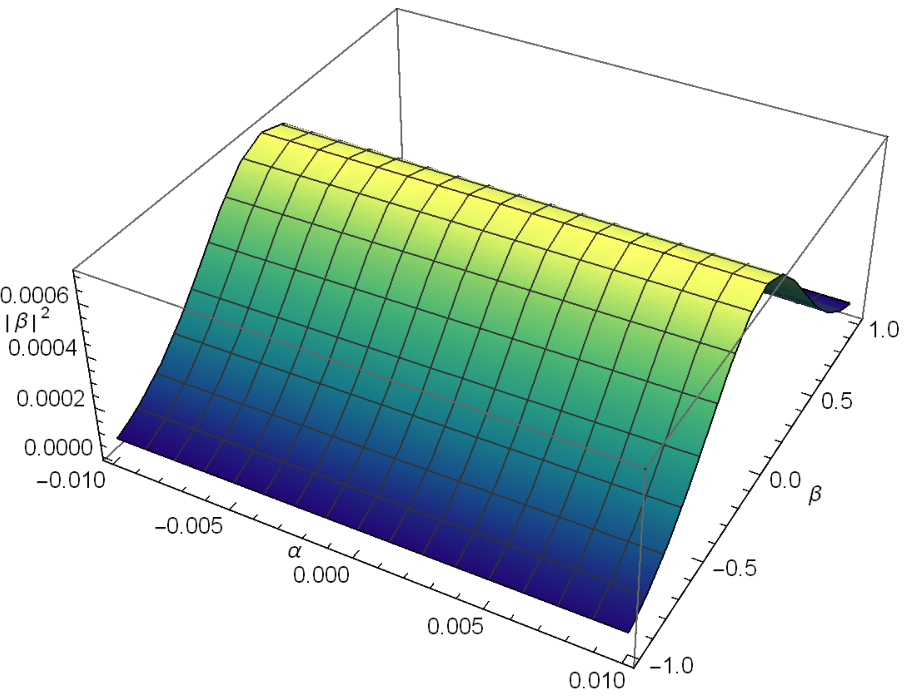}}\hfil
\subfloat[\{$\alpha, -1, 1$\}, \{$\beta, -0.01, 0.01$\}]
{\includegraphics[width=0.35\textwidth]{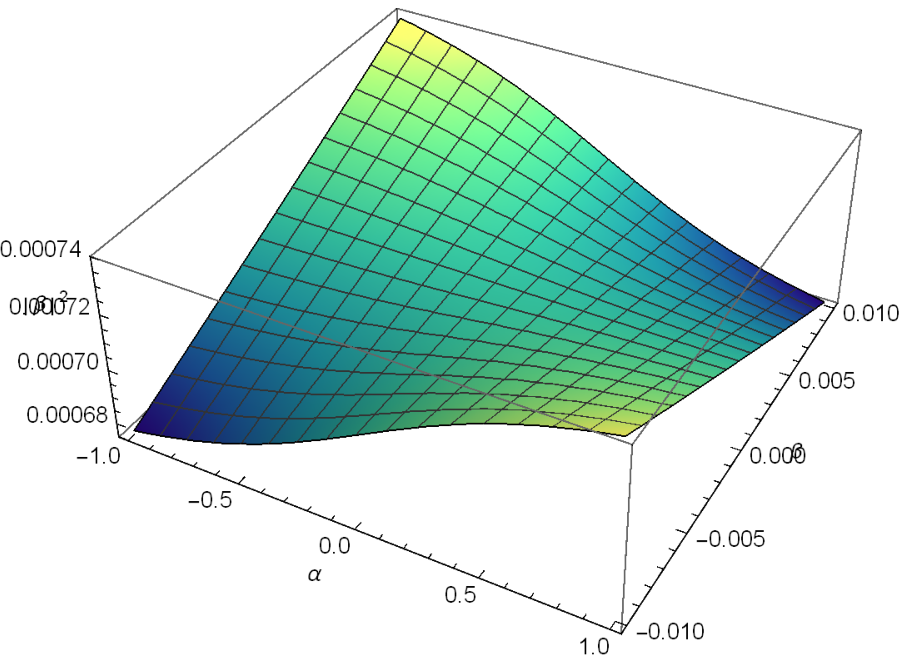}}
\caption{The Klein-Gordon particle number density for $m=1$. Here, $\alpha$ and $\beta$ are variables.}
\label{fig:example1}
\end{figure}
\FloatBarrier

It is also worth considering the phenomenon of weak and strong electric fields depending on the two parameters $\alpha$ and $\beta$, examining the behavior of the number density and deriving some relevant thermodynamic quantities. Firstly, let us assume that $\alpha=0$, $\beta\neq 0$, and the parameter $\lambda$ is zero. In this case, the created particle density \eqref{eq:61} takes the form
\begin{equation}\label{eq:62}
n_{k}=\frac{1}{\exp\left[2\pi\left(\sqrt{k_{\bot}^{2}+2\xi +e^{2}\beta^{2}}\right) \right] -1}.
\end{equation}
This density resembles a two-dimensional Bose-Einstein distribution and is thermal. For very weak electric fields, Eq. \eqref{eq:62} reduces to:
\begin{equation}\label{eq:63}
n_{k}=\frac{1}{\exp\left[2\pi\sqrt{k_{\bot }^{2}+2\xi }+\pi \frac{e^{2}\beta ^{2}}{\sqrt{k_{\bot }^{2}+2\xi }}\right] -1}\,.
\end{equation}

Equation \eqref{eq:61} indicates the contribution of weak electric fields to particle creation with a chemical potential proportional to $\frac{e^{2}\beta^{2}}{\sqrt{k_{\bot }^{2}+2\xi}}$. When $\beta\rightarrow 0$, the particle production \eqref{eq:63} becomes thermal
\begin{equation*}
n_{k}=\frac{1}{\exp \left[ 2\pi \sqrt{k_{\bot }^{2}+2\xi }\right] -1}\,.
\end{equation*}
In the absence of an electric field, this expression shows that the density of the created particles is a Bose-Einstein distribution, consistent with the results of references \cite{47,54}.

For extremely strong electric fields, Eq. \eqref{eq:62} reduces to
\begin{equation}\label{eq:64}
n_{k}\sim \exp \left[ -\pi \frac{k_{\bot }^{2}+2\xi }{e\beta }-2\pi e\beta \right].
\end{equation}
The formula \eqref{eq:64} represents a thermal distribution with a chemical potential proportional to $e\beta$. When $\beta \rightarrow \infty$, the particle production vanishes.

Secondly, let us assume that $\alpha \neq 0$ and $\beta = 0$, and the created particles density \eqref{eq:61} is of the form
\begin{equation}
n_{k}=\frac{1}{\exp \left[ 2\pi \sqrt{k_{\bot }^{2}+2\xi }\right] -1}
\end{equation}
This expression shows that the density of the created particles is a Bose-Einstein distribution, similar to the result obtained in the absence of an electric field \cite{54}.

From expression \eqref{eq:61}, it is evident that in the presence of the electric field \eqref{eq:12}, the distribution of created particles is not thermal, and it mainly depends on the influence of the electric field. However, if the electric field is proportional to the scalar curvature of space-time, then the created particles follow a Bose-Einstein distribution. It is noteworthy that when $m=0$, equation \eqref{eq:33} describes the massless scalar particle in the Bianchi I universe. In this limit, the number of created particles is determined by
\begin{equation}\label{eq:66}
n_{k}=\frac{\exp \left[ 2\pi \left( \sqrt{k_{\bot }^{2}+2\xi +e^{2}\beta ^{2}}-e\beta \right) \right] +1}{\exp \left[ 4\pi \left( \sqrt{k_{\bot}^{2}+2\xi +e^{2}\beta ^{2}}\right) \right] -1}\,.
\end{equation}
This is consistent with quantum field theory, where particles are created from the energy of the electromagnetic field.

The quantity $\varepsilon _{k}=\sqrt{k_{\bot }^{2}+2\xi +e^{2}\beta ^{2}}$ represents the energy. In the limit where $\varepsilon_{k}\gg e\beta$, the particle spectrum in equation \eqref{eq:66} approximates to
\begin{equation}
n_{k}=\frac{1}{\exp \left( 2\pi \varepsilon _{k}\right) -1}\sim \frac{1}{\exp \left( \frac{2\pi }{e\beta }\sqrt{k_{\bot }^{2}+2\xi }\right) -1}\,,
\end{equation}
This is equivalent to a Planckian distribution, where the temperature $T=e\beta /\left( 2\pi \right)$. The production of a thermal spectrum from a time-dependent electric field is possible under certain conditions. This provides an explanation for the formation of matter in the universe through the transformation of electromagnetic field energy into particles near the singularity. Experimental evidence, such as the direct creation of matter from photons and the Schwinger effect, supports the equivalence of energy and mass. These findings contribute to understanding the role of the electromagnetic field in the creation of matter and its profound influence on cosmic structures and particle formation.

\section{Dirac equation and particle creation process}

We solve the Dirac equation in the presence of the electric field \eqref{eq:12} within the cosmological background \eqref{eq:1} to determine the density of created Dirac particles. The simplest action for massive field spinors, when an electrodynamic gauge field $A_{\mu}$ is present in curved space-time, takes the form
\begin{equation}
\mathcal{S}=\frac{1}{2}\int d^{4}x\, e\left[i\left(\bar{\Psi}\tilde{D}\Psi-\overline{\tilde{D}\Psi}\Psi\right)-m\bar{\Psi}\Psi\right].
\end{equation}
The Dirac equation in the co-moving frame, utilizing the field equation \eqref{eq:field}, and denoting the generic field as $\bar{\Psi}$, is expressed as follows
\begin{equation}\label{eq:69}
\left(\tilde{D}-m\right)\Psi=0\,,
\end{equation}
with
\begin{equation}
\tilde{D}=\tilde{\gamma}^{\mu}\left(\partial_{\mu}-\Gamma_{\mu}-ieA_{\mu}\right),
\end{equation}
where the curved matrices $\tilde{\gamma}^{\mu}=e_{a}^{\mu}\gamma^{a}$, with $\gamma^{a}$ representing the Minkowski Gamma matrices, satisfy the Clifford algebra
\begin{equation}
\left\{ \tilde{\gamma}^{\mu},\tilde{\gamma}^{\nu}\right\}=2g^{\mu\nu}\,.
\end{equation}

Furthermore, the spin connection $\Gamma _{\mu }$ is represented as
\begin{equation}
\Gamma_{\mu}=\frac{1}{4}\tilde{\gamma}^{\nu}\left(\partial_{\mu}\tilde{\gamma}_{\nu}-\Gamma_{\mu\nu}^{\alpha}\tilde{\gamma}_{\alpha}\right).
\end{equation}
Then, the non-zero components of the spin connections are
\begin{equation}
\Gamma _{1}=\frac{1}{2}\gamma ^{0}\gamma ^{1}\,,\quad\Gamma _{2}=\frac{1}{2}\gamma^{0}\gamma ^{2}\,,\quad\Gamma _{3}=-\frac{1}{2t^{2}}\gamma ^{0}\gamma ^{3}\,.
\end{equation}
The form of the Dirac equation \eqref{eq:69} is
\begin{equation}\label{eq:74}
\left[\gamma ^{0}\partial_{0}+\frac{1}{t}\left(\gamma^{1}\partial_{1}+\gamma^{2}\partial_{2}\right)+\gamma^{3}\left(t\partial_{3}-ie\alpha-\frac{ie\beta}{t}\right)-m\right]\Phi=0\,,
\end{equation}
where
\begin{equation}
\Psi =t^{-1/2}\Phi\,.
\end{equation}

It is essential to point out that eq. \eqref{eq:74} may be represented as the sum of two first-order differential equations as follows
\begin{equation}\label{eq:76}
\left\{t\left[\gamma^{3}\partial_{0}+\gamma^{0}\left(t\partial_{3}-ie\alpha-\frac{ie\beta}{t}\right)-m\gamma^{3}\gamma^{0}\right]+\left(\gamma^{1}\partial_{1}+\gamma^{2}\partial_{2}\right)\gamma^{3}\gamma^{0}\right\}
\gamma^{3}\gamma^{0}\Phi=0\,.
\end{equation}
In the form of operators, this equation \eqref{eq:76} can be expressed as
\begin{equation}
\left\{K_{1}+K_{2}\right\}\phi =0\,,
\end{equation}
with
\begin{equation}\label{eq:78}
K_{2}\phi =k\phi =-K_{1}\phi\,,
\end{equation}
where
\begin{equation}
\phi=\gamma^{3}\gamma^{0}\Phi\,.
\end{equation}
The formulas for the operators $K_{1}$ and $K_{2}$ are as follows
\begin{align}
K_{1}&=t\left[\gamma ^{3}\partial _{0}+\gamma ^{0}\left( t\partial_{3}-ie\alpha -\frac{ie\beta }{t}\right) -m\gamma ^{3}\gamma ^{0}\right]\,, \\
K_{2}&=\left(\gamma ^{1}\partial _{1}+\gamma ^{2}\partial _{2}\right)\gamma ^{3}\gamma ^{0}\,,
\end{align}
where $k$ is the separation constant. It is possible to write the spinor $\phi$ as
\begin{equation}
\phi =\phi _{0}\exp \left( \vec{k}.\vec{r}\right),
\end{equation}
where $\phi_{0}$ is a bispinor as a function of $t$. We employ the representation in which the Gamma matrices in Minkowski space-time are defined by
\begin{equation}\label{eq:83}
\gamma^{0}=\left(\begin{array}{cc}
-i\sigma ^{1} & 0 \\
0 & i\sigma ^{1}
\end{array}\right),
\gamma ^{1}=\left(\begin{array}{cc}
0 & i \\
-i & 0
\end{array}\right),
\gamma ^{2}=\left(\begin{array}{cc}
\sigma ^{2} & 0 \\
0 & -\sigma ^{2}
\end{array}\right),
\gamma ^{3}=\left(\begin{array}{cc}
\sigma ^{3} & 0 \\
0 & -\sigma ^{3}
\end{array}\right).
\end{equation}

Since the $\sigma _{i}$'s represent the $2\times 2$ Pauli matrices, we conclude that equation \eqref{eq:78} simplifies to algebraic equations, enabling the identification of the relationship between the components of the bispinor $\phi _{0}$, as expressed by
\begin{equation}\label{eq:84}
\phi _{0}=\left(\begin{array}{c}
\phi _{1} \\
\phi _{2}
\end{array}\right) =\left(
\begin{array}{c}
\phi _{1} \\
\sigma _{2}\frac{k_{x}}{ik_{y}+k}\phi _{1}
\end{array}\right),
\end{equation}
where the eigenvalue $k$ is determined by
\begin{equation}
k=i\sqrt{k_{x}^{2}+k_{y}^{2}}=ik_{\perp}\,.
\end{equation}

We find that eq. \eqref{eq:76} can be simplified to the coupled system of first-order differential equations for $k_{z}=0$, by utilizing the representation in eq. \eqref{eq:83} and considering the spinor structure of eq. \eqref{eq:84}, where
\begin{align}
\left[ \sigma ^{3}\frac{d}{dt}+\sigma ^{1}\left( e\alpha +\frac{e\beta }{t}\right) -m\sigma ^{2}+\frac{k}{t}\right] \phi _{1} &=0\label{eq:86} \\
\left[ -\sigma ^{3}\frac{d}{dt}-\sigma ^{1}\left( e\alpha +\frac{e\beta }{t}\right) -m\sigma ^{2}+\frac{k}{t}\right] \phi _{2} &=0\label{eq:87}
\end{align}
It is easy to show that eq. \eqref{eq:87} is analogous to eq. \eqref{eq:86}. As a result, we have minimized the problem of solving eq. \eqref{eq:74} to finding the solution of eq. \eqref{eq:86}. Then, eq. \eqref{eq:86}, after multiplying it from the right by $\sigma ^{3}$, takes the following form
\begin{equation}\label{eq:88}
\left[\partial_{0}+i\sigma ^{2}\left( e\alpha -e\frac{\beta }{t}\right)+im\sigma ^{1}+\sigma ^{3}\frac{k}{t}\right] \Phi _{1}=0\,.
\end{equation}
We use the linear transformation $T$ \cite{55} to solve equation \eqref{eq:88}
\begin{equation}\label{eq:89}
T=\left(\begin{array}{cc}
a & -b \\
a & b
\end{array}\right),
\end{equation}
where $a$ and $b$ are nonzero arbitrary constants. After the application of the transformation $T$, equation \eqref{eq:88} is written as
\begin{equation}\label{eq:90}
\left[\partial _{0}+iT\sigma ^{2}T^{-1}\left( e\alpha -e\frac{\beta }{t}\right) +imT\sigma ^{1}T^{-1}+T\sigma ^{3}T^{-1}\frac{k}{t}\right] T\Phi_{1}=0\,,
\end{equation}
where the matrix transformation \eqref{eq:89} acts on the $\sigma ^{i}$ as follows
\begin{align}
T\sigma^{1}T^{-1}&=\frac{1}{2ab}\left(\begin{array}{cc}
-a^{2}-b^{2} & a^{2}-b^{2} \\
-a^{2}+b^{2} & a^{2}+b^{2}
\end{array}\right),\notag \\
iT\sigma^{2}T^{-1}&=\frac{1}{2ab}\left(\begin{array}{cc}
-a^{2}+b^{2} & a^{2}+b^{2} \\
-a^{2}-b^{2} & a^{2}-b^{2}
\end{array}\right),\label{eq:91}\\
T\sigma ^{3}T^{-1}&=\left(\begin{array}{cc}
0 & 1 \\
1 & 0
\end{array}\right)=\sigma^{1}\,,\notag
\end{align}
and
\begin{equation}\label{eq:92}
T\Phi=\digamma =\left(\begin{array}{c}
f_{1} \\
f_{2}
\end{array}\right).
\end{equation}

Now, substituting \eqref{eq:92} and \eqref{eq:91} into \eqref{eq:90}, we obtain the system of equations
\begin{align}
&\rho \frac{df_{1}}{d\rho }+\left[ \left( \frac{e\alpha }{4a^{2}b^{2}}\rho -\frac{e\beta }{2ab}\right) \left( -a^{2}+b^{2}\right) -im\frac{b^{2}+a^{2}}{4a^{2}b^{2}}\rho \right] f_{1}+\notag \\
&\left[ \left( \frac{e\alpha }{4a^{2}b^{2}}\rho -\frac{e\beta }{2ab}\right) \left( a^{2}+b^{2}\right)-im\frac{b^{2}-a^{2}}{4a^{2}b^{2}}\rho +k\right] f_{2}=0\,,\\
&\rho\frac{df_{2}}{d\rho }+\left[ -\left( \frac{e\alpha }{4a^{2}b^{2}}\rho -\frac{e\beta }{2ab}\right) \left( a^{2}+b^{2}\right) +im\frac{b^{2}-a^{2}}{4a^{2}b^{2}}\rho +k\right] f_{1}+  \notag \\
&\left[ \left( \frac{e\alpha }{4a^{2}b^{2}}\rho -\frac{e\beta }{2ab}\right) \left( a^{2}-b^{2}\right) +im\frac{b^{2}+a^{2}}{4a^{2}b^{2}}\rho\right] f_{2}=0\,,
\end{align}
where
\begin{equation}
\rho =2abt\,.
\end{equation}
For the simplest Whittaker equations, we choose $a$ and $b$ such that
\begin{align}
\frac{b^{2}-a^{2}}{2} &=-e\alpha\,,  \notag \\
\frac{b^{2}+a^{2}}{2} &=-im\,.
\end{align}
Then the system of equations becomes
\begin{align}
\rho \frac{df_{1}}{d\rho }+\left( \frac{1}{2}\rho +\frac{e^{2}\alpha \beta }{\sqrt{-m^{2}-e^{2}\alpha ^{2}}}\right) f_{1}+\left( \frac{ime\beta }{\sqrt{-m^{2}-e^{2}\alpha ^{2}}}+k\right) f_{2} &=0\,,  \notag \\
\rho \frac{df_{2}}{d\rho }-\left( \frac{1}{2}\rho +\frac{e^{2}\alpha \beta }{\sqrt{-m^{2}-e^{2}\alpha ^{2}}}\right) f_{2}-\left[ \frac{ime\beta }{\sqrt{-m^{2}-e^{2}\alpha ^{2}}}-k\right] f_{1} &=0\,.\label{eq:97}
\end{align}

By doing this, we have transformed the problem of solving eqs. \eqref{eq:88} into determining the solutions of eqs. \eqref{eq:97}. We find the second-order differential equation from eqs. \eqref{eq:97}
\begin{equation}\label{eq:98}
\frac{d^{2}f_{1}}{d\rho ^{2}}+\frac{1}{\rho }\frac{df_{1}}{d\rho }+\frac{\sigma }{\rho }f_{1}-\frac{\nu ^{2}}{\rho ^{2}}f_{1}-\frac{1}{4}f_{1}=0\,,
\end{equation}
where
\begin{equation}
\sigma =\frac{1}{2}+i\frac{e^{2}\alpha \beta }{\sqrt{m^{2}+e^{2}\alpha ^{2}}}\,,\qquad \nu =i\sqrt{k_{\perp }^{2}+e^{2}\beta ^{2}}\,.
\end{equation}
The wave function $f_{1}$ can be defined as
\begin{equation}
f_{1}=\rho ^{-1/2}\omega\,,
\end{equation}
and eq. \eqref{eq:98} is reduced to
\begin{equation}
\frac{d^{2}\omega _{1}}{d\rho ^{2}}+\left( \frac{\sigma }{\rho }+\frac{\frac{1}{4}-\nu ^{2}}{\rho ^{2}}-\frac{1}{4}\right) \omega =0\,.
\end{equation}

This equation is the Whittaker differential equation \cite{51,52,53,54}, therefore we can write the solution as a combination of Whittaker functions $M_{\tilde{k}_{\theta },\mu }\left(\rho \right)$ and $W_{\tilde{k}_{\theta} ,\mu}\left(\rho\right)$
\begin{equation}
\omega\left(\rho\right)=C_{1}\,M_{\sigma ,{\nu }}\left(\rho\right) +C_{2}W_{\sigma ,{\nu }}\left( \rho \right).
\end{equation}
The positive and negative frequency states as $t\rightarrow 0$ and $t\rightarrow \infty$ will be defined by the solutions asymptotic behavior
\begin{align}
f_{1}^{+}\left( t\rightarrow 0\right) &=C_{0}^{+}\rho ^{-1/2}M_{\sigma ,{\nu }}\left( \rho \right)\,, \\
f_{1}^{-}\left( t\rightarrow 0\right) &=\exp \left( \pi \left\vert {\nu }\right\vert \right) C_{0}^{-}\rho ^{-1/2}M_{\sigma ,-{\nu }}\left( \rho \right)\,,
\end{align}
and
\begin{align}
f_{1}^{+}\left( t\rightarrow \infty \right) &=C_{\infty }^{+}\rho^{-1/2}W_{\sigma ,{\nu }}\left( \rho \right), \\
f_{1}^{-}\left( t\rightarrow \infty \right) &=C_{\infty }^{-}\rho^{-1/2}W_{-\sigma ,{\nu }}\left( \rho \right),
\end{align}
where $C_{0}^{+}$, $C_{0}^{-}$, $C_{\infty }^{+}$, $C_{\infty }^{-}$ are normalisation constants. The Bogoliubov transformation can be used to calculate the density of created particles
\begin{equation}
f_{1}^{+}\left( t\rightarrow \infty \right) =\alpha f_{1}^{+}\left(t\rightarrow 0\right) +\beta f_{1}^{-}\left( t\rightarrow 0\right).
\end{equation}

Now, we use the relation
\begin{equation}
W_{\sigma ,{\nu }}\left( \rho \right) =\frac{\Gamma \left( -2{\nu }\right) }{\Gamma \left( \frac{1}{2}-\sigma -{\nu }\right) }M_{\sigma ,{\nu }}\left( \rho \right) +\frac{\Gamma \left( 2{\nu }\right) }{\Gamma \left( \frac{1}{2}-\sigma +{\nu }\right) }M_{\sigma ,-{\nu }}\left( \rho \right),
\end{equation}
in order to prove that $\alpha $ and $\beta $ are
\begin{equation}
\alpha =\frac{C_{\infty }^{+}\Gamma \left( -2{\nu }\right) }{C_{0}^{+}\Gamma \left( \frac{1}{2}-\sigma -{\nu }\right) }\,,\qquad
\beta =\exp \left( -\pi \left\vert {\nu }\right\vert \right) \frac{C_{\infty }^{+}\Gamma \left( 2{\nu }\right) }{C_{0}^{-}\Gamma \left(\frac{1}{2}-\sigma +{\nu }\right) }\,.
\end{equation}
Using the Gamma function's property
\begin{equation}
\left\vert \Gamma \left( ix\right) \right\vert ^{2}=\frac{1}{x\sin \pi x}\,,
\end{equation}
we obtain
\begin{equation}\label{eq:111}
\left\vert \frac{\beta }{\alpha }\right\vert ^{2}=\exp \left( -2\pi \sqrt{k_{\perp }^{2}+e^{2}\beta ^{2}}\right) \frac{\Theta _{-}\sinh \pi \Theta _{-}}{\Theta _{+}\sinh \pi \Theta _{+}}\,,
\end{equation}
where
\begin{equation}
\Theta _{\pm }=\sqrt{k_{\perp }^{2}+e^{2}\beta ^{2}}\pm \frac{e^{2}\alpha\beta }{\sqrt{m^{2}+e^{2}\alpha ^{2}}}\,.
\end{equation}
Using Eq. \eqref{eq:111} and the normalization condition for Dirac particles
\begin{equation}
\left\vert \alpha \right\vert ^{2}+\left\vert \beta \right\vert ^{2}=1\,,
\end{equation}
we obtain that the created Dirac particles density is
\begin{equation}\label{eq:114}
n_{k}^{D}=\left\vert \beta \right\vert ^{2}=\frac{\exp \left( -2\pi \sqrt{k_{\perp }^{2}+e^{2}\beta ^{2}}\right) \Theta _{-}\sinh \pi \Theta _{-}}{\exp \left( -2\pi \sqrt{k_{\perp }^{2}+e^{2}\beta ^{2}}\right) \Theta
_{-}\sinh \pi \Theta _{-}+\Theta _{+}\sinh \pi \Theta _{+}}\,.
\end{equation}

\begin{figure}[ht]
\centering
\subfloat[\{$\alpha, -0.01, 0.01$\},\{ $\beta, -1, 1$\}]
{\includegraphics[width=0.35\textwidth]{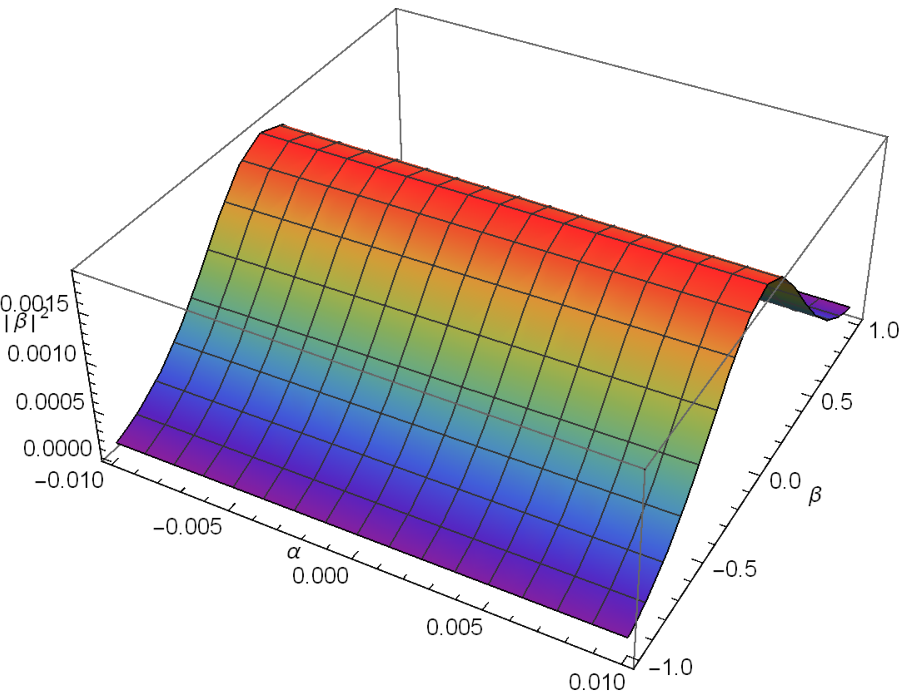}}\hfil
\subfloat[\{$\alpha, -1, 1$\}, \{$\beta, -0.01, 0.01$\}]
{\includegraphics[width=0.35\textwidth]{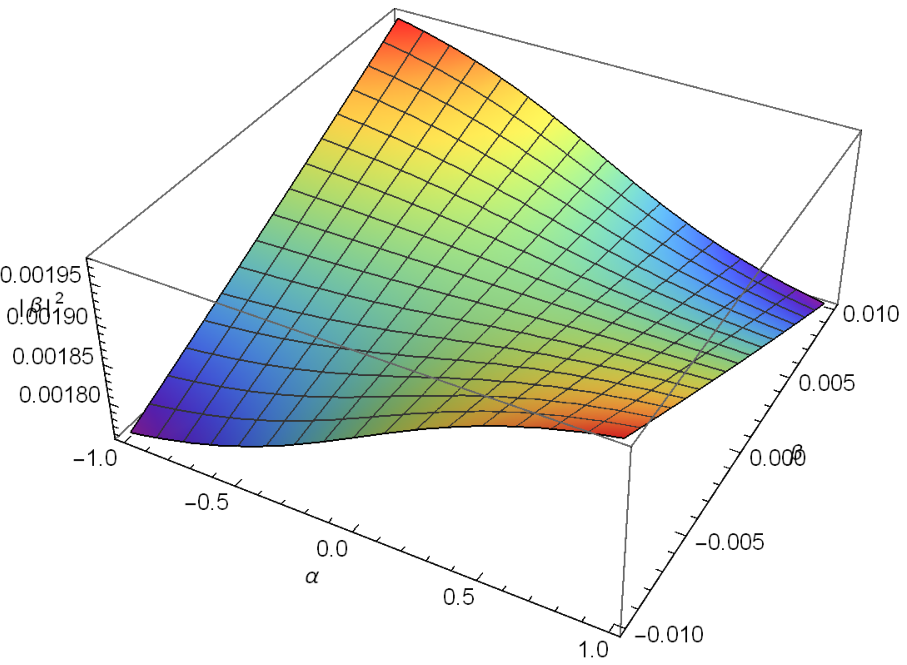}}
\caption{The Dirac particles number density when $m=1$. Here $\alpha$ and $\beta$ are variables.}
\label{fig:example2}
\end{figure}

Let us discuss the special cases:
\begin{description}
\item[Case $\boldsymbol{\alpha =0}$ and $\boldsymbol{\beta \neq 0}$:] In this case, the parameter$\Theta _{\pm }=\sqrt{k_{\perp }^{2}+e^{2}\beta ^{2}}$, and the created particles density \eqref{eq:114} has the following formula
\begin{equation}
n_{k}^{D}=\frac{1}{\exp \left[ 2\pi \left( \sqrt{k_{\bot }^{2}+e^{2}\beta^{2}}\right) \right] +1}\,.
\end{equation}
This density resembles a two-dimensional distribution of Fermi-Dirac and is thermal. The number density of created Dirac particles in the presence of strong electric fields takes the form
\begin{equation}\label{eq:116}
n_{k}^{D}\approx \exp \left[ -\pi \frac{k_{\bot }^{2}}{e\beta }-2\pi e\beta\right].
\end{equation}
The expression \eqref{eq:116} corresponds to a thermal distribution with a chemical potential proportional to $e\beta$.
\item[Case $\boldsymbol{\alpha \neq 0}$ and $\boldsymbol{\beta =0}$:] In this case, the created particles density (114) is of the form
\begin{equation}
n_{k}^{D}=\frac{1}{\exp \left( 2\pi k_{\bot }\right) +1}\,.
\end{equation}
This expression shows that the created particles density is a distribution of Fermi-Dirac, which is the same result as in the absence of an electric field \cite{55}.
\end{description}

Furthermore, if $m=0$, the equation \eqref{eq:74} describes a massless Dirac particle in the Bianchi I space-time. The formula for the number density of produced particles in this limit is given by
\begin{equation}\label{eq:118}
n_{k}^{D}=\frac{\exp \left( -2\pi \varepsilon _{k}\right) \mathring{\Theta}_{-}\sinh \pi \mathring{\Theta}_{-}}{\exp \left( -2\pi \varepsilon_{k}\right) \mathring{\Theta}_{-}\sinh \pi \mathring{\Theta}_{-}+
\mathring{\Theta}_{+}\sinh \pi \mathring{\Theta}_{+}}\,,
\end{equation}
where $\mathring{\Theta}=\varepsilon _{k}\pm e\beta $ and $\varepsilon_{k}=\sqrt{k{\bot }^{2}+e^{2}\beta^{2}}$. When we consider the limit $\varepsilon_{k}\gg e\beta $, the particle creation in equation \eqref{eq:118} can be approximated as
\begin{equation}
n_{k}^{D}=\frac{1}{\exp \left( 2\pi \varepsilon _{k}\right) +1}\sim \frac{1}{\exp \left( \frac{2\pi }{e\beta }k_{\bot }\right) +1}\,,
\end{equation}
which corresponds to a Fermi-Dirac distribution with temperature $T=e\beta/\left( 2\pi \right) $ and zero chemical potential. This result indicates that the universe expansion with a time-dependent electric field creates such particles thermally. Near the singularity, the transformation of electromagnetic field energy into particles is the origin of matter in the universe.

\section{Conclusions}

In this paper, the results obtained demonstrate that the quasi-classical approach enables the calculation of the density of created scalar and Dirac particles in a non-static space-time with a non-stationary electromagnetic source.

In Figures \ref{fig:example1} and \ref{fig:example2}, we illustrate the dependence of the number density of created particles on the strength of the electric field associated with the $\alpha$ and $\beta$ parameters. From the figures, it is evident that the individual contributions of the first term of the electric field, proportional to $t^{-1}$, and the second term, proportional to $t^{-2}$, to the particle creation density can be physically significant for electric fields proportional to $t^{-2}$. In this case, the particle density becomes subject to Bose-Einstein and Fermi-Dirac thermal distributions. If the second term of the electric field, proportional to $t^{-2}$, vanishes, the electric fields proportional to $t^{-1}$ do not contribute to the creation of particles influenced by the gravitational field.

Therefore, the particle distribution becomes thermal only under the influence of the electric field when the electric interaction is directly proportional to the Ricci scalar of curved space-time. We have also confirmed that the origin of the universe's matter is attributed to the transformation of electromagnetic field energy into particles near the singularity.

\section*{Acknowledgement}

This work is supported by PRFU research project B00L02UN050120190001, University of Batna-1, Algeria.

\end{document}